\documentclass{aa}  
\usepackage[colorlinks=true,citecolor=blue]{hyperref}
\usepackage{graphicx}
\usepackage{txfonts}
\begin{document} 

   \title{Modeling noise propagation in Fourier-filtering wavefront sensing, fundamental limits, and quantitative comparison}

   \author{V. Chambouleyron
          \inst{1}\fnmsep\inst{2}
          \and
          O. Fauvarque\inst{4}
          \and 
          C. Plantet\inst{5}
          \and
          J-F. Sauvage\inst{3}\fnmsep\inst{1}
          \and 
          N. Levraud\inst{1}\fnmsep\inst{3}
          \and
          M. Cissé\inst{1}\fnmsep\inst{3}
          \and
          B. Neichel\inst{1}
          \and
          T. Fusco\inst{3}\fnmsep\inst{1}
          }

   \institute{Aix Marseille Univ, CNRS, CNES, LAM, Marseille, France\\
              \email{vchambou@ucsc.edu}
         \and
             University of California Santa Cruz, 1156 High St, Santa Cruz, USA
         \and
             DOTA, ONERA, Université Paris Saclay, F-91123 Palaiseau, France
          \and
          IFREMER, Laboratoire Detection, Capteurs et Mesures (LDCM), Centre Bretagne, ZI de la Pointe du Diable, CS 10070, 29280, Plouzane, France
          \and
          INAF - Osservatorio Astronomico di Arcetri
             }

% \abstract{}{}{}{}{} 
% 5 {} token are mandatory
 
  \abstract
   {Adaptive optics (AO) is a technique allowing for ground-based telescopes' angular resolution to be improved drastically. The wavefront sensor (WFS) is one of the key components of such systems, driving the fundamental performance limitations.}
   { In this paper, we focus on a specific class of WFS: the Fourier-filtering wavefront sensors (FFWFSs). This class is known for its extremely high sensitivity. However, a clear and comprehensive noise propagation model for any kind of FFWFS is lacking.}
   {Considering read-out noise and photon noise, we derived a simple and comprehensive model allowing us to understand how these noises propagates in the phase reconstruction in the linear framework.}
   {This new noise propagation model works for any kind of FFWFS, and it allows one to revisit the fundamental sensitivity limit of these sensors. Furthermore, a new comparison between widely used FFWFSs is held. We focus on the two main FFWFS classes used: the Zernike WFS (ZWFS) and the pyramid WFS (PWFS), bringing new understanding of their behavior.}
   {}

   \keywords{wavefront sensing --
                Fourier-filtering wavefront sensor --
                noise propagation
               }

\maketitle

%------------------------------------------------------------------

\section{Introduction}

Adaptive optics (AO) is a technique allowing one to compensate for the impact of atmospheric turbulence on telescopes that has become essential for a large number of astrophysical applications. Motivated in particular by the hunt for exoplanets \citep{2012SPIE.8447E..0BK}, AO systems pushing the limits of performance are now being developed, and they are referred to as extreme AO systems (XAOs). These systems rely on high-order deformable mirrors with fast real-time computation. The fundamental limits of such instruments are based on the quality of the measurements provided by the optical device at the heart of this technique: the wavefront sensor (WFS). One key aspect driving the XAO instruments is the WFS sensitivity, which can limit the number of controlled modes and the speed of the loop. The Fourier-filtering WFS (FFWFS) represents a wide class of sensors, which are of particular interest thanks to their superior sensitivity. From a general point of view, a FFWFS consists of a phase mask located in an intermediate focal plane which performs an optical Fourier filtering. This filtering operation allows for the conversion of phase information at the entrance pupil into amplitude at the pupil plane, where a quadratic sensor is used to record the signal \citep{2004OptCo.233...27V, Guyon_2005, fauvOptica}. The FFWFS can exhibit an extra optical stage called modulation, which consists in making the electromagnetic field spin around the tip of the filtering mask during one integration time of the detector. This well-known operation allows for the dynamic range to be increased at the expense of the sensitivity \citep{raga}.

The goal of this paper is to introduce a simple and comprehensive model for noise propagation through any FFWFS, allowing one to quantitatively compare them within a fixed and clear framework. Inspired by an approach already employed for the Shack-Hartmann wavefront sensor \citep{1992Rigaut}, we first derived a noise propagation model in the linear regime that can be applied to any FFWFS (section 2). We then show all the potential of this model by interpreting the behavior of two widely used classes of FFWFSs in a new way in section 3: the Zernike WFS (ZWFS) and the pyramid WFS (PWFS).

\section{Noise propagation and sensitivity in the linear regime}

\subsection{Reduced intensities: A standard way to process FFWFS signal}

For this study, the FFWFS signal has been processed in a standard way leading to a quantity, referred to as reduced intensities. If we called $I(\phi)$ the intensities recorded on the detector for a given phase $\phi$, we built the reduced intensities $\Delta I(\phi)$ the following way:

\begin{equation}
    \Delta I(\phi) = \frac{I(\phi)}{N_{ph}}-I_{0}
,,\end{equation}

where $N_{ph}$ is the number of photo-electrons available for the measurement and $I_{0}$ is the reference intensities. We would like to specify that the normalization by the total flux is absolutely required; otherwise, a change in flux would be interpreted as phase aberration. In this paper, we assume that we are working around a flat wavefront and we therefore define the reference intensities $I_{0}$ as the normalized intensities recorded on the WFS detector for a flat wavefront in the pupil:

\begin{equation}
    I_{0} = \frac{I(\phi=0)}{N_{ph}}
.\end{equation}

The reduced intensities provide the simplest way to process FFWFS signal by taking in account all the pixels on the detector, which is a procedure independent from the shape of the mask and structure of the modulation \citep{fauva2017}. 

\subsection{Noise propagation in the linear regime}

To derive the noise propagation model, we made some assumptions: (i) we made the small-phase approximation, which allowed us to assume that the FFWFSs are working in their linear regime; and (ii)  we also assumed that we were working with a monochromatic light of wavelength $\lambda$. In that framework, the FFWFS response to a modal basis $[\phi_{1}, ..., \phi_{i}, ..., \phi_{N}]$ (for which the rms value of each mode is set to 1 in any chosen unit) is fully characterized by the interaction matrix $\mathcal{D}$ built through a push-pull calibration process:

\begin{equation}
    \mathcal{D} = [\delta I(\phi_{1}),..., \delta I(\phi_{i}),..., \delta I(\phi_{N})]
,\end{equation}

where $\delta I(\phi_{i})$ is constructed the following way:
\begin{equation}
    \delta I(\phi_{i}) = \frac{\Delta I(\epsilon\phi_{i}) - \Delta I(-\epsilon\phi_{i})}{2\epsilon }
,\end{equation}

and where $\epsilon$ is a scalar small enough to remain in the linear regime of the sensor. By computing the reconstructor $\mathcal{D}^{\dag}$ defined as the pseudo-inverse of $\mathcal{D}$ ({i.e.,} $\mathcal{D}^{\dag}=(\mathcal{D}^{t}\mathcal{D})^{-1}\mathcal{D}^{t}$) and assuming small-phase approximation, we have the following relationship, which allow us to reconstruct the incoming phase, provided that the sensing conditions were the same as calibration (same wavelength $\lambda$, same pupil, etc.):

\begin{equation}
\phi= \mathcal{D}^{\dag}\Delta I(\phi)\\
.\end{equation}

When the signal that is delivered by the camera is affected by noise, one can write the following:

\begin{equation}
    I_{b}(\phi) = I(\phi) + b(\phi)
,\end{equation}

where $b$ is the noise, which may depend on the phase to be measured. By applying the reduced intensities' computation on these noisy intensities (assuming a noise-free $I_{0}$), one gets the following:

\begin{equation}
    \Delta I_{b}(\phi) = \Delta I(\phi) + \frac{b(\phi)}{N_{ph}} = \Delta I(\phi) + n(\phi)
,\end{equation}

where $n(\phi)$ can be seen as a "noise-to-signal ratio." The smaller its value is, the less reduced intensities are affected by noise. This quantity propagates in the phase reconstruction in the following way:

\begin{equation}
\begin{split}
\phi+\xi &= \mathcal{D}^{\dag}(\Delta I(\phi)+n(\phi))\\
\xi &= \mathcal{D}^{\dag}n(\phi)
\end{split}
,\end{equation}

where $\xi$ is therefore the phase estimation error due to noise. One can study the statistical behavior of this phase error with respect to the statistical behavior of noise. For that, we write $<.>$ which is the operator averaging on noise realizations. We can compute the error covariance matrix as follows:

\begin{equation}
<\xi\xi^{t}> = \mathcal{D}^{\dag}<n(\phi)n^{t}(\phi)>\mathcal{D}^{\dag t}
\label{noise_propagation}
,\end{equation}

where $<n(\phi)n^{t}(\phi)>$ is the noise covariance matrix \citep{1992Rigaut}. Starting from this equation, it is possible to establish a noise propagation model for all FFWFS sensors.

\subsubsection{Sensitivity to read-out noise}

We first focus on the behavior with respect to the read-out noise. This noise only depends on the detector; it is therefore independent from the incoming phase and the FFWFS characteristics (namely the mask and the modulation). Assuming that each pixel noise is decorrelated with the others and that noise is the same for each pixel with a standard deviation $\sigma_{ron}$ (in $e^{-}$/px/frame), one can write the following:

\begin{equation}
\label{eq:bruit_uniforme_propagation}
<n(\phi)n^{t}(\phi)>=<nn^{t}> = \frac{\sigma_{ron}^{2}}{N_{ph}^{2}}\mathbf{I}
,\end{equation}

where $\mathbf{I}$ is the identity matrix. Equation \ref{noise_propagation} can be rewritten as follows:

\begin{equation}
<\xi\xi^{t}> = \frac{\sigma_{ron}^{2}}{N_{ph}^{2}}\mathcal{D}^{\dag}\mathcal{D}^{\dag t}= \frac{\sigma_{ron}^{2}}{N_{ph}^{2}}(\mathcal{D}^{t}\mathcal{D})^{-1}
\label{RON}
.\end{equation}

By using the modal basis generated by the singular-value decomposition of $\mathcal{D}$, $\mathcal{D}^{t}\mathcal{D}$ becomes diagonal. We assume this assertion to be still true for any modal basis. This assumption is quite strong because it is stating (to some extend) that there is no cross talk between measurements, which is not always the case for the considered FFWFSs. One should carefully analyze the elements of ($\mathcal{D}^{t}\mathcal{D})^{-1}$ to fully grasp the inversion process at stake \citep{1992Rigaut}. Nonetheless, assuming $\mathcal{D}^{t}\mathcal{D}$ as being diagonal allows one to  drastically simplify the study because it is then possible to compute the variance of each mode due to noise propagation with the diagonal terms of equation \ref{RON}. For a given mode $\phi_{i}$, it leads to the following:

\begin{equation}
\label{varaince_mode_i}
\sigma^{2}_{\phi_{i}} = \frac{\sigma_{ron}^{2}}{(\mathcal{D}^{t}\mathcal{D})_{i,i}\times N_{ph}^{2}}
,\end{equation}

which gives, in standard deviation,
\begin{equation}
\label{ecart_mode_i}
\sigma_{\phi_{i}} = \frac{\sqrt{N_{sap}}\times\sigma_{ron}}{\sqrt{N_{sap}}\times\sqrt{(\mathcal{D}^{t}\mathcal{D})_{i,i}}\times N_{ph}}
.\end{equation}

In this equation, we have included $N_{sap}$, which is the number of subapertures used for the measurement in the pupil. It allows for the sensitivity $s(\phi_{i})$ to read-out noise for a given mode $\phi_{i}$ to be defined as follows:

\begin{equation}
s(\phi_{i}) = \sqrt{N_{sap}}\times\sqrt{(\mathcal{D}^{t}\mathcal{D})_{i,i}}
\label{sensi_uniform}
.\end{equation}

Adding the quantity $\sqrt{N_{sap}}$ allows for different masks for any given sampling to be compared, provided that the considered mode $\phi_{i}$ is well sampled. Equation \ref{sensi_uniform} can be reformulated in the following way: 
\begin{equation}
\label{sensibilité_mode_i}
s(\phi_{i}) = \sqrt{N_{sap}}\times\big|\big|\delta I(\phi_{i})\big|\big|_{2}
,\end{equation}

with the sensitivity to RON for a given mode $\phi_{i}$ therefore being the two-norm of its corresponding column in the interaction matrix. One can understand from equation \ref{ecart_mode_i} that a higher sensitivity subsequently leads to less noise propagation. Therefore, we aim to design WFS reaching the higher sensitivity possible.

\subsubsection{Sensitivity to photon noise}

Photon noise follows a Poisson distribution, and therefore the variance of the distribution is equal to the average of the intensities. Hence, we can assume that $<n(\phi)n^{t}(\phi)>$ is diagonal and write the following:

\begin{equation}
\label{bruit_photon}
<n(\phi)n^{t}(\phi)> = \frac{\textbf{diag}(I(\phi))}{N_{ph}^{2}}
,\end{equation}

where $\textbf{diag}(I(\phi))$ is a square matrix whose diagonal elements are the values of $I(\phi)$. However, by definition, this noise depends on the illumination pattern on the detector. It therefore depends on the modulation, the mask, and also the incoming phase. Because we are working in the linear regime, we can assume that

\begin{equation}
    I(\phi) = I_{0} + \Delta I(\phi) \approx I_{0},\ \text{for }\ \phi<<1
,\end{equation}

which means that the intensities on the detector can be approximated by the ones corresponding to a flat wavefront. Because $I_{0}$ is already a normalized map, $\textbf{diag}(I(\phi))=N_{ph}\times\textbf{diag}(I_{0})$. We can then write 

\begin{equation}
\label{bruit_photon2}
<n(\phi)n^{t}(\phi)> = <nn^{t}> = \frac{\textbf{diag}(I_{0})}{N_{ph}}
,\end{equation}

and by injecting this formula in equation \ref{noise_propagation},

\begin{equation}
\label{eq:bruit_photon_propa}
<\xi\xi^{t}>_{\tau} = \frac{1}{N_{ph}}\mathcal{D}^{\dag}\textbf{diag}(I_{0})\mathcal{D}^{\dag t}
.\end{equation}

Since $\mathcal{D}^{\dag}\mathcal{D}^{\dag t} = (\mathcal{D}^{t}\mathcal{D})^{-1}
$, one can rearrange the previous equation as follows:

\begin{equation}
<\xi\xi^{t}>_{\tau} = \frac{1}{N_{ph}}\big((\textbf{diag}(1/\sqrt{I_{0}})\mathcal{D})^{t}\textbf{diag}(1/\sqrt{I_{0}})\mathcal{D}\big)^{-1}
.\end{equation}

As was done for the RON noise, assuming the diagonality of the matrix
\begin{equation}
\nonumber
(\textbf{diag}(1/\sqrt{I_{0}})\mathcal{D})^{t}\textbf{diag}(1/\sqrt{I_{0}})\mathcal{D}
\end{equation}

allows onne to simply write the phase error for a given mode $\phi_{i}$ as follows:

\begin{equation}
\label{ecart_mode_i_photons}
\sigma_{\phi_{i}} = \frac{1}{\sqrt{[(\textbf{diag}(1/\sqrt{I_{0}})\mathcal{D})^{t}\textbf{diag}(1/\sqrt{I_{0}})\mathcal{D}]_{i,i}} \times \sqrt{N}}
.\end{equation}

Due to this last formula, we decided to call sensitivity to photon noise $s_{\gamma}$ the following quantity:

\begin{equation}
s_{\gamma}(\phi_{i}) = \sqrt{[(\textbf{diag}(1/\sqrt{I_{0}})\mathcal{D})^{t}\textbf{diag}(1/\sqrt{I_{0}})\mathcal{D}]_{i,i}}
\label{eq:ch3:sensi_photon}
.\end{equation}

In a more comprehensive way, the photon noise sensitivity to the mode $\phi_{i}$ is calculated by the following operation:

\begin{equation}
\label{sensibilité_mode_i_photon}
s_{\gamma}(\phi_{i}) = \big|\big|\frac{\delta I(\phi_{i})}{\sqrt{I_{0}}}\big|\big|_{2}
.\end{equation}

One could wonder if there should be a special procedure for nonilluminated pixels in $I_{0}$. In that case, these pixels exhibit a null value in the interaction matrix and add no contribution to the sensitivity computation. The formulas given in \ref{ecart_mode_i} and \ref{ecart_mode_i_photons} give a relationship between the phase error, the number of photons available for the measurement, the RON, the number of subapertures, and the intrinsic properties of the studied FFWFSs. The noise propagation for a given mode $\phi_{i}$ is then quantified through the formula

\begin{equation}
\boxed{
\sigma^{2}_{\phi_{i}} = \frac{N_{sap}\times\sigma_{ron}^{2}}{ s^{2}(\phi_{i})\times N_{ph}^{2}}+\frac{1}{s^{2}_{\gamma}(\phi_{i})\times N_{ph}}}
\label{eq:ch4:propagation_bruit_mesure}
,\end{equation}

where $\sigma^{2}_{\phi_{i}}$ is given in $u^{2}$, and where $u$ is the unit used to normalize the modes to rms 1 when building the interaction matrix (usually radians or nanometers). We would like to mention that $s(\phi_{i})$ and $s_{\gamma}(\phi_{i})$ have no reason to be equal. When comparing different FFWFSs in terms of sensitivity, we thus need to compare their RON sensitivity and their photon noise sensitivity. This kind of formula was already derived in \citet{Guyon_2005} for simplified and idealized cases of 4PWFS and ZWFS. Providing their interaction matrix and the linear regime, formula \ref{eq:ch4:propagation_bruit_mesure} provides a unified scheme that works for any kind of modulation and mask, and it does not use an analytical model: the computed sensitivities are therefore more accurate with this derivation.

\subsection{Maximum sensitivity limit}

The columns of the interaction matrix are encoding the derivative of the signal with respect to the phase. Hence, the sensitivity expressions given in equations \ref{sensibilité_mode_i} and \ref{sensibilité_mode_i_photon} are actually describing a metric called Fisher information \citep{2015OExpr..2328619P}. This metric is bounded by the Cramer-Rao bound, and it is then possible to show that both sensitivities defined here (for RON and photon noise) have a maximum value of 2 \citep{paterson,PhysRevApplied.15.024047}:

\begin{equation}
\begin{split}
    0\leq\ &s \leq 2\\
    0\leq\ &s_{\gamma} \leq 2
\end{split}
.\end{equation}

These boundaries are therefore useful to provide an absolute reference to compare FFWFS sensitivities. We would like to add that the Cramer-Rao bound actually works for any kind of WFS, and these boundaries extend out of the FFWFS scope.

\section{Sensitivity comparison of different FFWFS}

The goal of this section is to use the metrics defined previously to compare well-known and widely used FFWFSs. We subsequently focus on two main classes of FFWSs: ZWFS and PWFS. We subsequently show that the use of the RON sensitivity, on the one hand, and the photon noise sensitivity, on the other hand, highlight some important properties of these sensors. For our comparison, we computed sensitivities with respect to the Fourier modes, which are simply defined by the sum of a cosine and a sine carrying a given spatial frequency $\textbf{f}$ \citep{fauvConv}. The following quantity then encodes the sensitivity
\begin{equation}
s_{\textbf{f}} = \frac{1}{\sqrt{2}}\sqrt{s\Big(\cos_{\textbf{f}}\Big)^{2}+s\Big(\sin_{\textbf{f}}\Big)^{2}}\,.
\label{eq:e2e_fourier}
\end{equation}

\subsection{Zernike class}

The ZWFS is known for its extremely high sensitivity while having a low dynamic range, which makes it a unique sensor for second stage AO systems or quasi-static aberrations' calibration sensor \citep{jensen-clem,ZeldaMamadou,Ruane_2020}. This sensor is composed of a focal plane mask made of a phase shifting dot which can be fully described by two parameters: its diameter $p$ and phase shift $d$. In a previous paper \citep{2021A&A...650L...8C}, we extensively described and analyzed the impact that the influence of the mask parameters have on sensitivity, but only for RON. We have extended this study by applying our metric developed for photon noise sensitivity to have a better understanding of the behavior of this sensor. Usually, the ZWFS is used in what we call here the classical ZWFS configuration: $p=1.06\ \lambda/D$ and $d=\pi/2$ \citep{viganSphere}. We remind the reader that the linear signal for the ZWFS is fully located in the footprint of the pupil image onto the detector plane, meaning that all photons outside of the footprint are left unused in the linear regime \citep{fauva2017}.

\subsubsection{Diameter}

\begin{figure}[h!]
    \centering
    \includegraphics[scale=0.8]{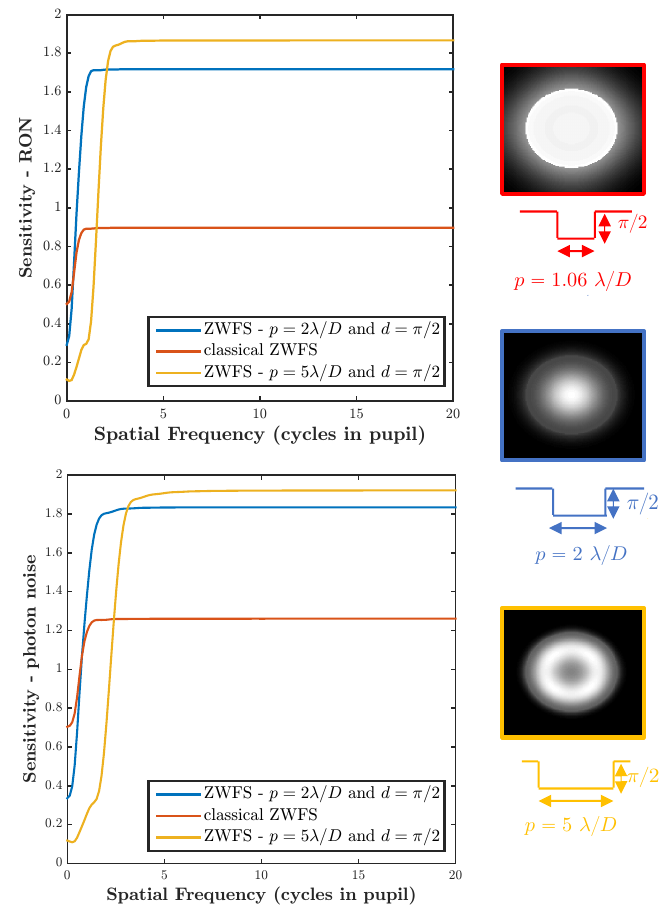}
    \caption{Senstivities to RON and photon noise for a different configuration of the ZWFS. Phase shift fixed at $d = \pi/2$ and diameter $p=1.06\ \lambda/D$, $p=2\ \lambda/D$, and $p=5\ \lambda/D$. We can see that the two kind of sensitivities follow the same trends.}
    \label{fig:ZWFS_diameter}
\end{figure}

It was shown that increasing the ZWFS diameter can significantly increase the RON sensitivity for high spatial frequencies, at the expense of low spatial frequency located inside the dot perimeter \citep{2021A&A...650L...8C}. Our goal is to assess if this statement also applies for the photon noise sensitivity. In this study, we have fixed the dot phase shift to $\pi/2$. Figure \ref{fig:ZWFS_diameter} shows the sensitivity for RON and photon noise computed in the case of three different diameters: $p=1.06\ \lambda/D$, $p=2\ \lambda/D$, and $p=5\ \lambda/D$. We see that the behavior for RON described in \cite{2021A&A...650L...8C} is also seen for photon noise: a high sensitivity gain for high spatial frequencies and a drop in sensitivity for frequencies located inside the dot. We would like to remind readers that changing mask parameters also affects the reference intensities $I_{0}$, which are plotted in figure \ref{fig:ZWFS_diameter} for the considered configurations. Finally, all conclusions that were drawn in the Z2WFS study are still valid for the photon noise sensitivity.

\subsubsection{Phase shift}

\begin{figure}[h!]
    \centering
    \includegraphics[scale=0.5]{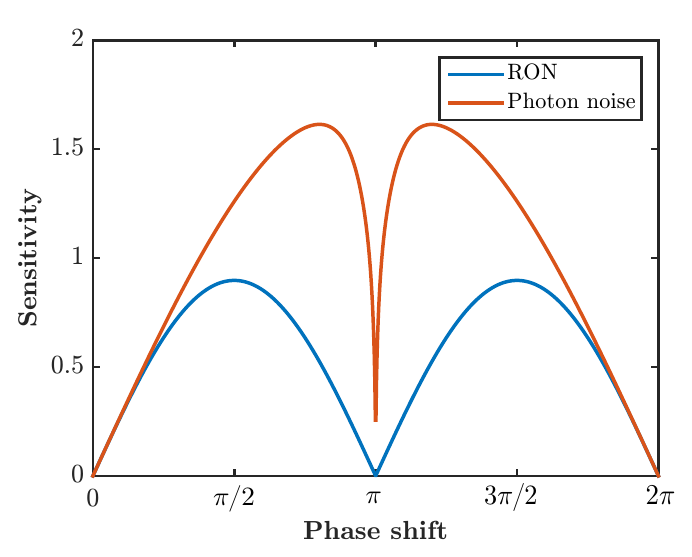}
    \caption{Sensitivities for a given spatial frequency corresponding to six cycles in the diameter for $p = 1.06\ \lambda/D$, while varying phase shift $d$. The maximum values are reached for a different value of $d$.}
    \label{fig:ZWFS_depthVariation_1}
\end{figure}

We now focus on the impact of the dot phase shift on sensitivity. It is already know that for RON sensitivity, the maximum is reached for $d = \pi/2$ no matter what is chosen as diameter size $p$. One could wonder if this assertion remains true for photon noise sensitivity. As a first step to answer this question, we fixed $p = 1.06 \lambda/D$ and let the depth vary between 0 and $2\pi$. We focused on the sensitivity for a given spatial frequency corresponding to six cycles in the diameter. The choice of this spatial frequency is arbitrary, but motivated by the fact that its footprint in the focal plane is located far from the phase shifting dot boundaries and not inside the phase-shifting area. The ZWFS behavior with respect to this frequency is then representative of all the high-order modes. The results are given Figure \ref{fig:ZWFS_depthVariation_1}. We can see that for the RON sensitivity, the maximum value is indeed reached at $d=\pi/2$. We also notice the coronagraphic configuration corresponding to $d = \pi$ for which the intensities on the detector do not have any linear dependence on the entrance phase \citep{Roddier_1997}. One can also notice the symmetric behavior of this curve, which is perfectly logical because of the $2\pi$ wrapping of the phase. For the photon noise sensitivity, the curve exhibits a slightly different behavior: the maximum value is not reached for $\pi/2$, but for $d=0.8\pi$. This configuration is very instructive because it gives us profound insight into how optimal RON sensitivity configurations can differ from optimal photon noise sensitivity configurations. 

\begin{figure}[h!]
    \centering
    \includegraphics[scale=0.5]{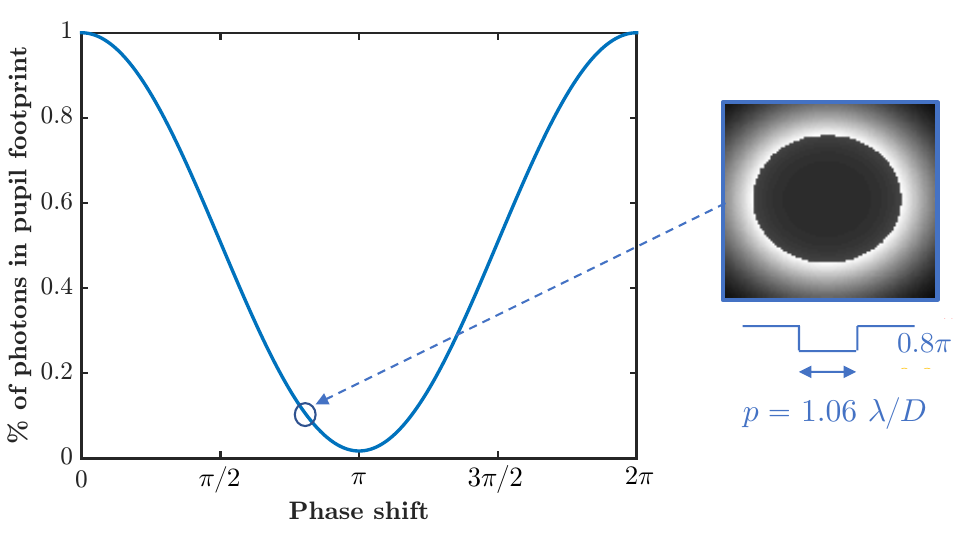}
    \caption{Percentage of photons kept in the pupil footprint for a configuration with diameter $p = 1.06 \lambda/D$, while changing phase shift $d$. The reference intensities for $d = 0.8\pi$ are represented on the right.}
    \label{fig:ZWFS_depthVariation_2}
\end{figure}

To understand why $d=0.8\pi$ incites better performance with respect to photon noise, we have to focus on the structure of the reference intensities. In the case of a WFS sensitivity for photon noise, photons are the source of signal but also of noise. We have shown the importance of the reference intensities in the previous section (Equation \ref{eq:ch3:sensi_photon}). In the configuration $p = 1.06 \lambda/D$ and $d=0.8\pi$, the sensor exhibits a behavior close to the coronagraphic one. Therefore, the reference intensities have a limited number of photons in the pupil footprint, as is shown in Figure \ref{fig:ZWFS_depthVariation_2} which shows the percentage of photons available in the pupil footprint of the reference intensities while increasing phase shift for diameter $p = 1.06 \lambda/D$. Here is a way to understand the photon noise sensitivity curve's shape that is presented in figure \ref{fig:ZWFS_depthVariation_1}: in assuming uniform reference intensities for the configuration $p = 1.06 \lambda/D$ while changing the depth $d$, it can be interpreted as the ratio between the RON sensitivity curve given in the same figure and the pupil footprint flux curve shown in figure \ref{fig:ZWFS_depthVariation_2}. The maximum found at $d=0.8\pi$ can therefore be seen as a trade-off between high signal ($d=\pi/2$) and low noise produced by photons ($d=\pi$). This configuration is a nice example of the "dark-WFS" concept, which is a class of sensor for which reference intensities would gather a limited number of photons \citep{darkWFS}.

\begin{figure}[h!]
    \centering
    \includegraphics[scale=0.8]{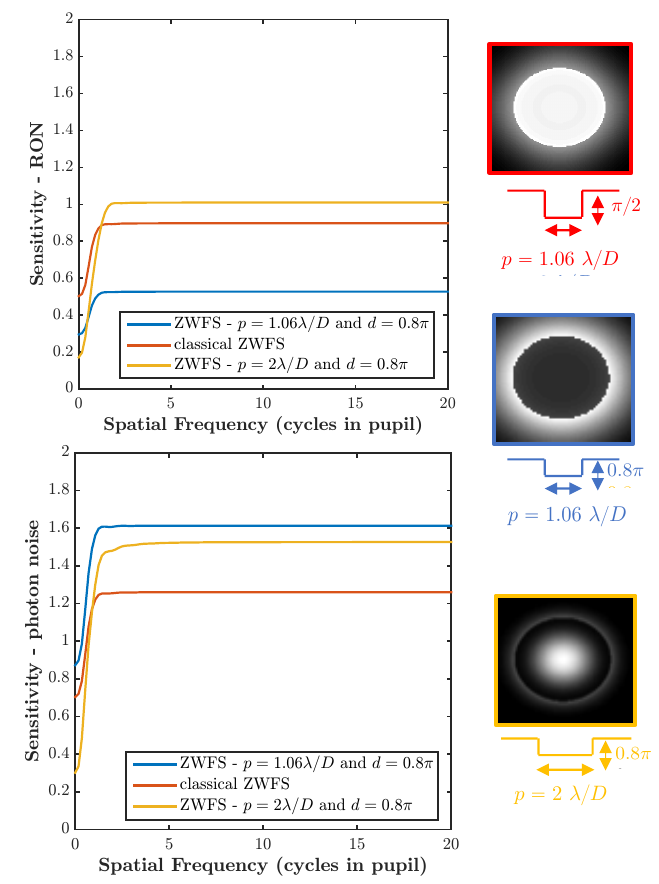}
    \caption{Sensitivities to RON and photon noise for a different configuration of the ZWFS. The "dark" ZWFS configuration corresponds to $p = 1.06 \lambda/D$ and $d=0.8\pi$.}
    \label{fig:ZWFS_depth}
\end{figure}

We then plotted the sensitivity for all the spatial frequencies for different configurations, as was done in the previous subsection when varying the diameter $p$. Here, we continue to explore the instructive $d=0.8\pi$ configuration as an example by comparing the classical ZWFS configurations with two other sets of parameters for the dot shape: $p = 1.06 \lambda/D$ with $d=0.8\pi$, and $p = 2 \lambda/D$ with $d=0.8\pi$. The results are given in Figure \ref{fig:ZWFS_depth}. For the  $p = 1.06 \lambda/D$ with a $d=0.8\pi$ configuration, the RON sensitivity is lower for all spatial frequencies compared to the classical ZWFS, while it is always higher for photon noise sensitivity. However, increasing the dot diameter to a value of $p = 2 \lambda/D$ while having $d=0.8\pi$ decreases the gain for photon noise sensitivity for all frequencies. This can be easily explained by the fact that in this configuration, the large size of the dot prevents coronagraphic behavior and does not remove a significant amount of photons from the pupil footprint in reference intensities. 

One could push the analysis  further by studying the photon noise sensitivity behavior for a full set of different diameter $p$ and phase shift $d$ and for all spatial frequencies. That is out of the scope of this paper. We have demonstrated the benefit of using two different sensitivity metrics for RON and photon noise and have therefore introduced the required tools to carry out a fine study of the ZWFS class.

\subsection{Pyramid class}

In this section, we study another FFWFS class: the PWFS. This class relies on a pyramidal shape of the phase mask, which splits the electromagnetic field into different beams \citep{raga}. For the PWFS, the extra optical module called "modulation" presented in the introduction is added. We study the impact of the following parameters on sensitivities: (i) the number of faces, which determines the number of pupil images that formed on the detector; and (ii) the angle of the faces, which determines the relative separation between pupil images.

\subsubsection{Number of faces}

\begin{figure}[h!]
    \centering
    \includegraphics[scale=0.35]{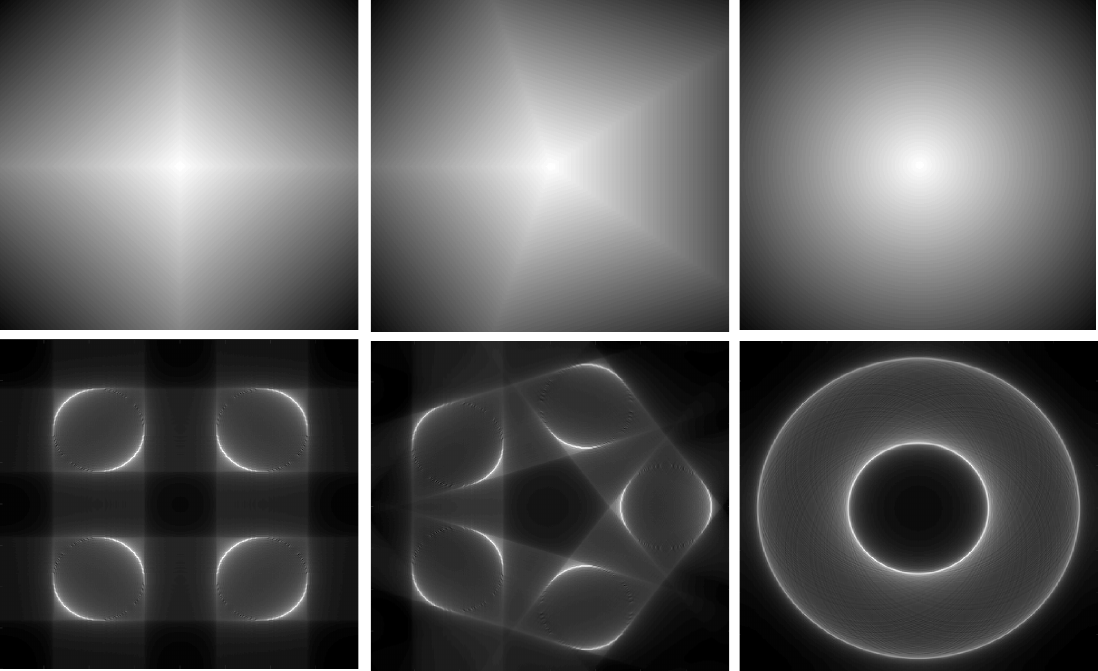}
    \caption{PWFS with a different number of faces. \textbf{Top:} arg(mask) for 4PWFS, 5PWFS, and axicon. \textbf{Botton:} Reference intensities associated to the nonmodulated case.}
    \label{fig:PWFS_faces_im}
\end{figure}

We start this study by analyzing the impact of the number of faces on PWFS sensitivities. To that end, we use five different configurations: a three-sided pyramid (3PWFS), the most widely used one; a four-sided pyramid (4PWFS); a five-sided pyramid (5PWFS); a six-sided pyramid (6PWFS); and finally a pyramid with an infinite number of faces, forming what we call an axicon. As an example, in Figure \ref{fig:PWFS_faces_im}, we show the phase masks and the corresponding reference intensities of the 4PWFS, 5PWFS, and axicon in the nonmodulated case.

We assess the impact of the number of faces on sensitivities by considering two modulation cases: a modulated case with a given modulation radius $r_{mod} = 3\ \lambda/D$ and the nonmodulated case $r_{mod} = 0\ \lambda/D$. Because the considered masks are not axisymmetric (except for axicon), we could analyze their sensitivities in the frequencies' space by extracting 2D sensitivity maps. For an easier comparison, we decided to extract radial cuts from these maps. For that, we chose a direction which does not lie on any edges for the considered configurations. \\

\begin{figure}[h!]
    \centering
    \includegraphics[scale=0.4]{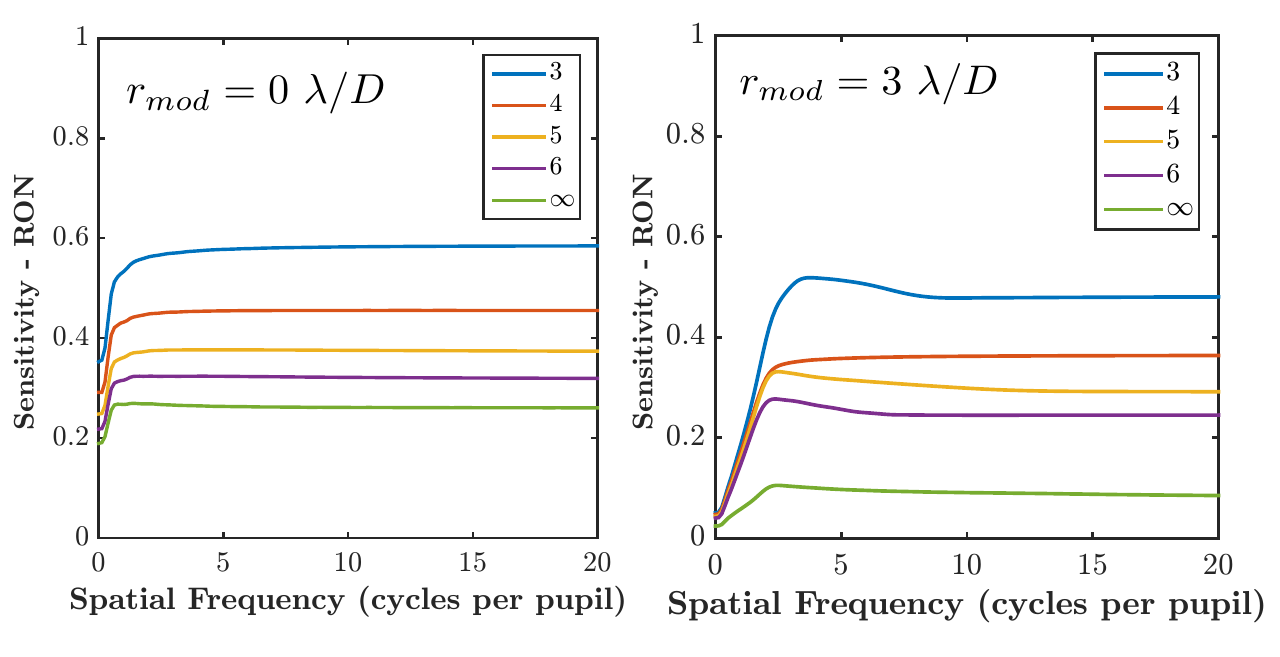}
    \caption{RON sensitivities for the PWFS when changing the number of faces. \textbf{Left:} Nonmodulated cases. \textbf{Right:} Modulation at $3\ \lambda/D$.}
    \label{fig:PWFS_faces}
\end{figure}

\textit{RON sensitivity:} We analyze first the RON sensitivity. Figure \ref{fig:PWFS_faces} shows the results of our study for different PWFS cases. We would like to point out that the sensitivity scale stops at 1 for better readability, but we remind readers that the maximum sensitivity is 2. 

In the nonmodulated case, we note that reducing the number of faces increases the sensitivity. The gain in sensitivity is logically proportional to the number of useful pixels: if the linear signal reaches more pixels, the PWFS lets more RON propagate. Therefore, the plateau for the 3PWFS is four-thirds higher than the 4PWFS, for example. When modulating (right panel Figure \ref{fig:PWFS_faces}), we observed the same relative behavior: the fewer faces, the better. We can see the well-known loss of sensitivity for spatial frequencies located in the modulation radius. We also spotted a global loss for all spatial frequencies. One can also note the different behavior for the axicon: its sensitivity is decreasing drastically compared to other configurations. \\

\begin{figure}[h!]
    \centering
    \includegraphics[scale=0.4]{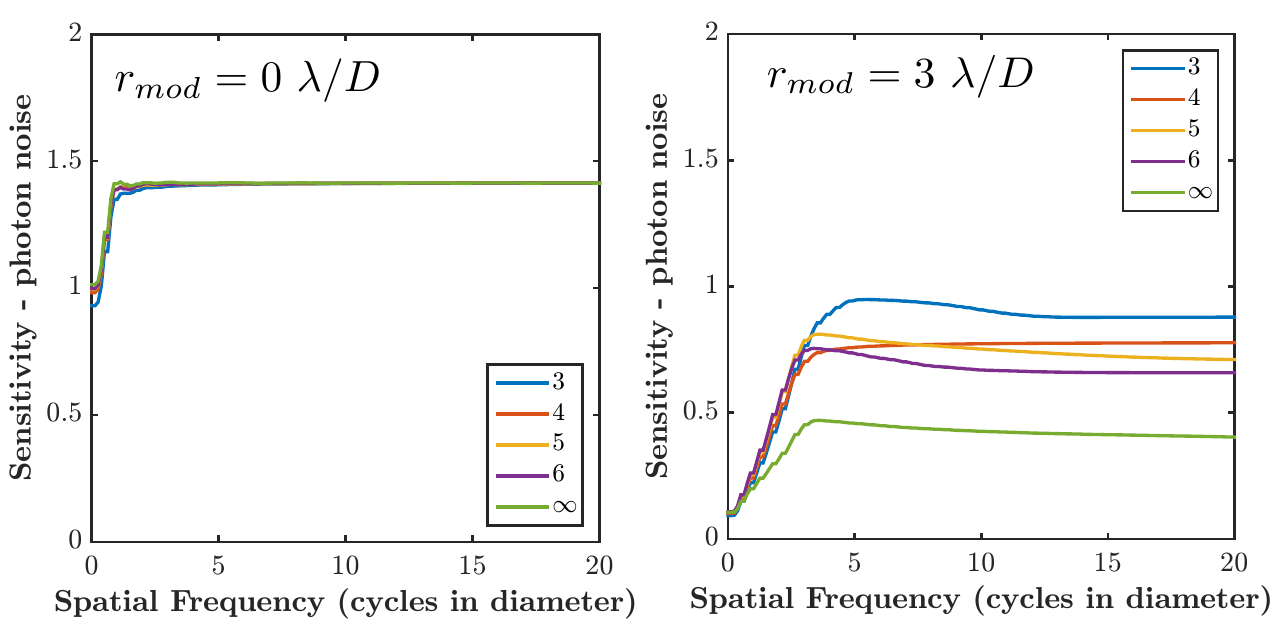}
    \caption{Photon noise sensitivities for the PWFS when changing the number of faces. \textbf{Left:} Nonmodulated cases. \textbf{Right:} Modulation at $3\ \lambda/D$.}
    \label{fig:PWFS_faces_photons}
\end{figure}

\textit{Photon noise sensitivity:} We now analyze the impact of the number of faces with respect to photon noise sensitivity. The results are given in Figure \ref{fig:PWFS_faces_photons}. We start our analysis with the modulated case (right panel). As for the RON sensitivity, we see that the sensitivity increases when reducing the number of faces. When modulating, the distribution of photons is mainly uniform in the pupil images. The ratio between sensitivities is then proportional to the square root of the number of faces ratio because the signal-to-noise ratio increases with the square root of the number of photons. For instance, the 3PWFS exhibits an increased sensitivity of $\sqrt{4}/\sqrt{3}$ with respect to the 4PWFS. However, when looking at the nonmodulated cases (Figure \ref{fig:PWFS_faces_photons}, left panel), all the different PWFS configurations exhibit the same sensitivities. This behavior can be explained by the fact that, in the nonmodulated case, diffraction effects by the edges of the masks have a significant impact on the intensities recorded on the detector.

This study shows that the 3PWFS exhibits a better sensitivity than the other configurations for the RON sensitivity (as expected and already tackled in previous studies \citep{2021arXiv210906386S}). We show here for the first time that this superiority is also true for photon noise sensitivity, but only when modulated.

We also emphasize a key aspect when comparing PWFS to ZWFS: the nonmodulated PWFSs have a better photon noise sensitivity than the classical ZWFS ($p=1.06\ \lambda/D$ and $d = \pi/2$). This observation seems important since the classical ZWFS is often presented as the most sensitive FFWFS. We see here that in its classical configuration, that is far from being the case because its sensitivities do not reach the maximum value of 2.

\subsubsection{Angle}

\begin{figure}[h!]
    \centering
    \includegraphics[scale=0.4]{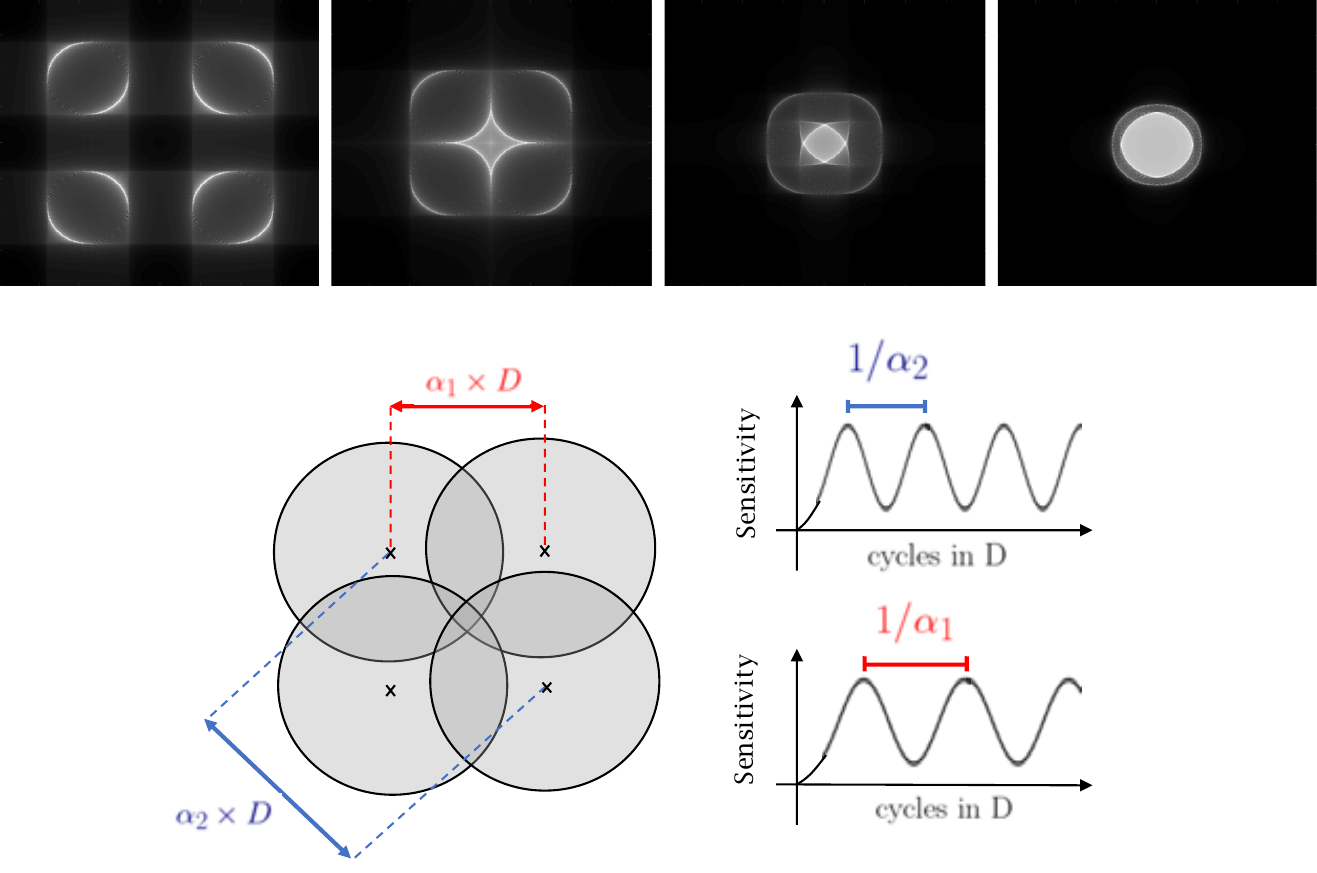}
    \caption{Flattened PWFS: periodicity of the oscillating sensitivity depends on the pupil overlapping ratio.}
    \label{fig:PWFS_flatten}
\end{figure}

We now study the impact of the pyramid angle on sensitivities. For this study, we chose a nonmodulated 4PWFS. We already know that decreasing the PWFS angle can change its behavior as soon as the pupils are overlapping, creating a configuration called the flattened PWFS (FPWFS) which can be interpreted as a mix between a PWFS and a shearing-interferometer. We call a classical PWFS the one showing no overlapping. It was already demonstrated \citep{fauvFP} that the FPWFS class exhibits an oscillating sensitivity curve, whose period depends on the separation between pupil images (illustration given in figure \ref{fig:PWFS_flatten}).

\begin{figure}[h!]
    \centering
    \includegraphics[scale=0.7]{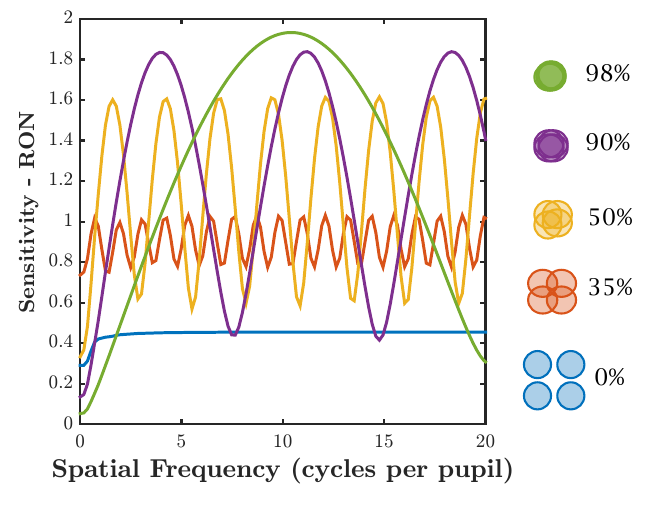}
    \caption{RON sensitivities for different FPWFS configurations (nonmodulated cases).}
    \label{fig:PWFS_angles}
\end{figure}

\textit{RON sensitivity:} Following the same procedure as before, we started this study with the sensitivity to RON. This aspect of the FPWFS has already been studied. We used this example of five configurations with a different overlapping ratio (percentage of pupil diameter), and plotted their respective sensitivity curves in figure \ref{fig:PWFS_angles}. We find the well-know oscillating curves. One can notice that there are some configurations that show a better sensitivity to RON with respect to the classical PWFS for all spatial frequencies. However, one could wonder if such configurations are still more sensitive than the classical PWFS when considering the photon noise sensitivity. Thanks to our new propagation model, we can assess this question for the first time.\\

\begin{figure}[h!]
    \centering
    \includegraphics[scale=0.7]{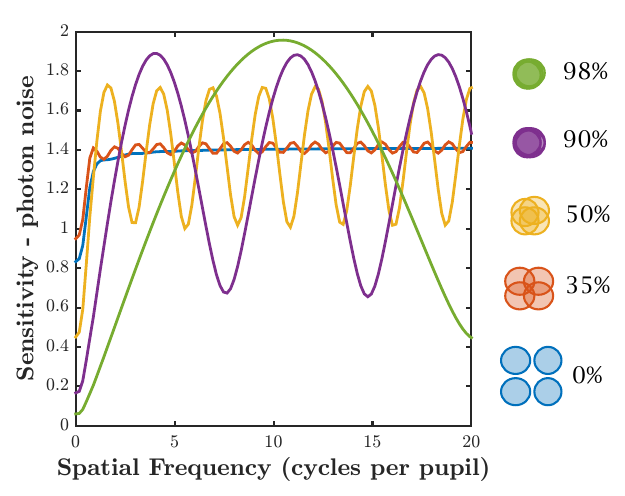}
    \caption{Photon noise sensitivities for different FPWFS configurations (nonmodulated cases).}
    \label{fig:PWFS_angles_photon}
\end{figure}

\textit{Photon noise sensitivity:} We computed the sensitivity to photon noise with the same overlapping configurations studied just before. Sensitivity curves are given figure \ref{fig:PWFS_angles_photon}. These sensitivity curves show a very different story: this time, we can see that there is no configuration which is always above the classical PWFS. If some configuration has a better sensitivity for some frequencies, it is always at the expense of others. The curves are oscillating around the classical PWFS one. These results bring a new understanding of the FPWFS class, showing that the FPWFS configurations are not as interesting as one could have expected for the photon noise sensitivity. The sensitivity curves given in figures \ref{fig:PWFS_angles} and \ref{fig:PWFS_angles_photon} can be compared to the ones computed for the ZWFS class given in figure \ref{fig:ZWFS_depthVariation_1}, where one can notice that in its optimal configurations, the ZWFS class outreaches the PWFS for a large set of spatial frequencies.

\section{Conclusion and perspectives}

In this paper, we have derived a simple and comprehensive noise propagation model for all FFWFSs. This model relies on the small phases' approximation and allows one to define two kinds of sensitivities which describe FFWFS behavior with respect to noise. There is the RON sensitivity which is computed through the interaction matrix, and the photon noise sensitivity which is computed through the interaction matrix and the reference intensities. These two bounded quantities can be used to link the number of photons available for the phase measurement and the estimation error variance. It provides a unified model which allows one to quantitatively compare all FFWFSs and that goes beyond the results given in \cite{Guyon_2005}. We have demonstrated the usefulness of this model by applying it to ZWFS and PWFS, highlighting for the first time some interesting properties of these sensors and exhibiting unprecedented comparisons. The fact that we now have well-defined criteria for the RON and photon noise sensitivities can lead us to try to optimize the mask shape to reach sensitivities close to the maximum value of 2 \citep{chambou_SPIE2022}.

This noise propagation model was derived assuming a linear regime. Here, we did not intend to analyze the dynamic range of the studied FFWFSs. Therefore, this study is just giving a part of the full description these sensors. On top of that, it is well known that FFWFSs are not working in their linear regime when working in closed-loop, leading to nonlinear effects that can be described at first order by a loss of sensitivity for each mode (the so-called optical gain effect \cite{Chambouleyron2020}). One could therefore define the effective sensitivities, which would be the sensitivities defined in this paper, weighted by the loss of sensitivities induced by nonlinearities.

%These noise propagation model was derived assuming linear regime. We did not intend here to analyze dynamic range of the studied FFWFS. Therefore this study is just giving a part of the full description these sensors. On top of that, it is well known that FFWFS are not working in their linear regime when working in closed-loop, leading to non-linear effects than can be described by optical gains: a loss of sensitivity for each mode. One could therefore define the effective sensitivities, which would be the sensitivities defined in this paper, weighted by the optical gains. 

\section{acknowledgements}
This work benefited from the support of the WOLF project ANR-18-CE31-0018 of the French National Research Agency (ANR). It has also been prepared as part of the activities of OPTICON H2020 (2017-2020) Work Package 1 (Calibration and test tools for AO assisted E-ELT instruments). OPTICON is supported by the Horizon 2020 Framework Programme of  the  European  Commission’s  (Grant  number  730890). Authors are acknowledging the support by the Action Spécifique Haute Résolution Angulaire (ASHRA) of CNRS/INSU co-funded by CNES. Vincent Chambouleyron PhD is co-funded by "Région Sud" and ONERA, in collaboration with First Light Imaging. Finally, part of this work is supported by the LabEx FOCUS ANR-11-LABX-0013.

% Bibliography
\bibliographystyle{aa} % style aa.bst
\bibliography{biblio}

\end{document}